\documentclass{aa}
\input psfig.sty
\begin{document}
\title{A 1.4~GHz radio continuum and polarization survey at medium
Galactic latitudes:}
\subtitle{II. First section}
\author{B.~Uyan{\i}ker, E.~F\"urst, W.~Reich, P.~Reich, and R.~Wielebinski}
\institute{Max--Planck--Institut f\"ur Radioastronomie,
Postfach 2024, 53010 Bonn, Germany}
\thesaurus{20(04.19.1; 09.19.1; 10.07.1; 13.18.3) }
\offprints{W.~Reich}
\date{Received 30 September 1998 / Accepted 8 April 1999}
\titlerunning{A 1.4~GHz survey at medium Galactic latitudes II.}
\authorrunning{B. Uyan{\i}ker et al.}

\maketitle

\begin{abstract}
We present the first section of a radio continuum and
polarization survey at medium Galactic latitudes carried out with the
Effelsberg 100-m telescope at 1.4~GHz. Four large fields have been
observed, which all together cover an area of about $1100~\sq$.
The rms-sensitivity is about 15~mK~T$_\mathrm{B}$ (about 7~mJy/beam
area) in total intensity and is limited by confusion. A sensitivity of
8~mK T$_\mathrm{B}$ is obtained in linear polarization. The angular
resolution of the observations is 9\farcm35. The maps in total
intensity and linear polarization have been absolutely calibrated by
low resolution data where available. Significant linear polarization is
seen in all the maps. In general, the intensity fluctuations measured
in linear polarization are not correlated with total intensity
structures. Areas of high polarization of some degrees extent are
seen, again with no apparent corresponding total intensity feature.
Modulation of polarized background emission by spatially varying
Faraday rotation seems the most likely explanation. Quite unexpected
is the detection of filamentary and ring-like depolarization
structures in the direction of the anticentre region, whose extents 
are up to about $3\degr$.

\end{abstract}

\section{Introduction} \label{intoduction}

Our knowledge of the Galactic magnetic field at present is primarily
based on studies of the distribution of rotation measures of pulsars or
extragalactic sources and optical observations of polarized star light
(e.g. Wielebinski\ \cite{wielebinski92}, Beck et al.\ \cite{beck+96}).
In addition, early radio polarization surveys at low frequencies have
revealed substantial linear polarization from the diffuse Galactic
emission, implying the general presence of ordered magnetic fields.
However, these early results were limited by low angular resolution and
sensitivity. Also Faraday rotation effects are significant at low
frequencies.

A polarimetric survey of the northern Galactic plane 
has been made by Junkes et al. (\cite{junkes+87a}) and Duncan et al.\
(\cite{duncan+99}) at 2695~MHz with the Effelsberg 100-m telescope.
Complementary observations have been made with the Parkes 64-m telescope
of a strip of the southern Galactic plane at 2400~MHz by Duncan et al.
(\cite{duncan+97}).
These observations revealed polarized emission of the diffuse
Galactic radiation. At least some of these emission features are
believed to originate at distances up to a few kpc (Junkes et al.\
\cite{junkes+87b}). The medium and high latitudes of the Galaxy
have not been studied
in a systematic way. Surprisingly, observations at 327~MHz by Wieringa et al.
(\cite{wieringa+93}) carried out with the Westerbork radio telescope
revealed filamentary polarized structures on a degree
scale. These high Galactic latitude features are not
detected at higher frequencies and have no counterpart in total
intensity. Recently, Gray et al. (\cite{gray+98}) detected
at 1.4~GHz an interstellar Faraday rotation feature
about $2\degr$ in size located in front of the prominent \ion{H}{ii}
region W5. These results call for a systematic observational approach.

There exist already a Galactic plane survey at 1.4~GHz
(Reich et al.\ \cite{reich+90a},\ \cite{reich+97}) carried out with the
Effelsberg telescope, which covers absolute Galactic latitudes of $4\degr$.
However, this survey does not include measurements of linearly
polarized intensities. Meanwhile, a new sensitive receiver
 is available and we have started a sensitive survey of the radio continuum
and polarized emission at medium Galactic latitudes with the Effelsberg
telescope. 
 
In a first paper (Uyan{\i}ker et al.\ 1998,\ \cite{paper1}) we have
described the methods of observation and data reduction.
In addition, we explained
the procedure of absolute calibration of both the total intensity data
from the Stockert 1.4~GHz survey (Reich\ \cite{reich82}, Reich \& Reich\
\cite{reich+reich86}) and the linear polarization data by 1.4~GHz
observations made with the Dwingeloo 25-m telescope (Brouw \& Spoelstra\
\cite{brouw+spoelstra76}). The medium latitude Effelsberg
survey aims to observe  regions at Galactic latitudes ($\vert b
\vert \leq 20\degr$) accessible to the telescope.

The areas presented in this paper cover four fields. From the first
Galactic quadrant an area of $10\degr \times 16\degr$ centered on
$\ell = 50\degr,\ b = 12\degr$ has been observed. Here the Dwingeloo
polarization data could not be used because of a low S/N-ratio in
general. The Cygnus-X area is the second region. This field includes
the Cygnus superbubble, which is particularly bright in X-ray emission.
The third area is the highly polarized region just above the Galactic
plane between $140\degr \leq \ell \leq 153\degr$. In the anticentre a
field of $20\degr \times 11\fdg 2$ centered on  $\ell = 200\degr$,
$b = 9\fdg 4$ was surveyed. For this region no Dwingeloo polarization
data are available for an absolute calibration.

In Sect.~2 we summarize some technical aspects.
In Sect.~3 we show that the total intensity data are
limited by confusion and we compare counts of compact sources with
those based on the recent 1.4~GHz NVSS survey. The total intensity and
polarimetric observations are displayed in Sect.~4, where in addition 
some comments on the individual fields are given.

\section{The Survey: first section} \label{surv}

The observations have been made with the two channel HEMT 1.3--1.7~GHz
receiver installed in the  primary-focus  of the Effelsberg 100-m
telescope. The method of observation is to scan each field along
Galactic latitude and longitude. Both coverages are
combined as described in \cite{paper1}. The half-power-beam-width at
1.4~GHz is $9\farcm 35 \pm 0\farcm 04$. The
conversion factor from main beam brightness temperature T$_\mathrm{B}$
to flux density/beam area is $2.12\pm 0.02$.
Details of the data reduction are given in \cite{paper1}, where we also
describe the method to reduce the instrumental polarization to a level of
about 1\%. The total intensity maps have been calibrated to an absolute
scale using the Stockert 1.4~GHz survey and the polarization data by 
Dwingeloo 1.4~GHz data.  Unfortunately, the Dwingeloo survey is incomplete.
For some areas the data are severely undersampled or a have a low S/N-ratio.

\section{Sensitivity, compact sources and confusion limit} \label{conf}

All total intensity survey maps show a large number of compact
sources, which are mainly of extragalactic origin. We have selected
several small flat and empty fields from the survey maps and subtracted
sources when necessary. The measured rms-noise is typically about
15~mK~T$_\mathrm{B}$. No problem with confusing sources exists for
polarized intensities, where a typical rms-noise value of 
8~mK~T$_\mathrm{B}$ has been found.

\begin{figure}[htb] 
\psfig{file=fig1.ps,width=8.8cm,bbllx=60pt,bblly=105pt,bburx=460pt,bbury=510pt,clip=}
\caption[]{Source counts from an area in the Galactic anticentre as
described in Sect.~3.}
\label{scall}
\end{figure}

The measured rms-noise in total intensity is about a factor of three
larger than that calculated from the system temperature of about
30~K. Source confusion is believed to be the limiting factor in
sensitivity. Condon et al.
(\cite{condon+89}) have studied the effect of source confusion for the
former 300-ft Green-Bank telescope, which is described by
$\mathrm{\Delta S_c} = 50\, \nu^{-2.7}$, where $\nu$ is in GHz and
$\mathrm{\Delta S_c}$ is the rms of intensity fluctuations in units of
mJy/300-ft beam area. When calculating the expected confusion for the
100-m telescope from that approach we get about 16~mJy or
35~mK~T$_\mathrm{B}$ at 1.4~GHz, which is significantly larger than the
measured rms-noise of 15~mK~T$_\mathrm{B}$ in the maps. We conclude
that source confusion is significantly smaller than previously thought,
but it is limiting the sensitivity for total intensities of the survey.

We have performed source counts from the survey maps based on a fitted
two-dimensional Gaussian to each source. These counts have been
compared with counts based on the recent 1.4~GHz VLA-survey (NVSS) by
Condon et al. (\cite{condon+98}), which is more sensitive to compact
sources than our survey and suffers less from confusion due to its
smaller beam size. We convolved a VLA map covering about $71.5~\sq$ to
the angular resolution of the Effelsberg map and applied the same source
fitting procedure to both maps. In Fig.~\ref{scall} we show cumulative
source counts for a region in the Galactic anticentre. We have also
fitted the VLA data at their original angular resolution and found
910 sources in total. The corresponding cumulative source counts are also
shown in Fig.~\ref{scall} with a slope of about $-1.4$. This is
close to the expected value of $-1.5$ for a uniform source density in
the local universe. Source evolution and different source populations
cause deviations from the $-1.5$ slope for sources weaker than about
100~mJy (see Condon et al.\ \cite{condon+98} for details). Below
about 40~mJy the fraction of sources which can not be fitted
individually increases with decreasing flux density. For the
Effelsberg data as well as for the convolved VLA data we obtain the
same result over the entire flux density range. However, the deviation
from the straight line starts already near 100~mJy. This effect is
caused by the much larger confusion in the low resolution data: The
total number of sources stronger than 10~mJy found in the original VLA
data results in a mean source separation of about 22~VLA beam areas,
but just about 1.8~beam areas of the 100-m telescope. Therefore, flux
densities lower than about 100~mJy source counts are more complete,
when using the original VLA data. With decreasing flux density an
increasing fraction of sources cannot be fitted individually due to
increasing confusion effects.

\section{Results}   \label{results}

We present the results of our observations in the form of contour and
grey-scale maps. The maps are also available in FITS-format via
Internet ({\tt http://www.mpifr-bonn.mpg.de/survey.html}).

When adding the absolutely calibrated large-scale emission to the
Effelsberg data  in both total and polarized intensity, much of the
small-scale details are hidden in large-scale intensity gradients. We
therefore separated the small and large-scale structures from each
other using the "background filtering method" (Sofue \& Reich\
\cite{sofue+reich79}), which is based on an unsharp masking operation.
This procedure has been applied to all absolutely calibrated
total intensity maps.
We separated emission on scales larger than about
$3\degr$ from compact sources and emission on smaller scales. The sum
of both components is exactly the original intensity. This procedure
and presentation of data has been already applied for the Effelsberg
Galactic plane surveys at 1.4~GHz (Reich et al.\ \cite{reich+90a},
\cite{reich+97}) and 2.695~MHz (Reich et al.\ \cite{reich+90b},
F\"urst et al.\ \cite{furst+90}).

\begin{figure*}[h] 
\hfill{}
\psfig{file=fig2.ps,width=13.5cm,%
bbllx=80pt,bblly=75pt,bburx=510pt,bbury=760pt,clip=}
\hfill{}
\caption[]{Small-scale total intensity image of the region towards $\ell
= 50\degr$ with superimposed polarization  vectors in the E-field direction.
Galactic coordinates are shown. The first contour set starts from
0~mK~T$_\mathrm{B}$ and runs in steps of 120~mK~T$_\mathrm{B}$ and the
second contour set starts at 600~mK~T$_\mathrm{B}$ and runs in steps of
240~mK~T$_\mathrm{B}$. The wedge at the top shows the lower and upper cuts
of the image. Every second polarization vector is plotted and a vector
of $2\arcmin$ length corresponds to 100~mK~T$_\mathrm{B}$ in polarized
intensity.}
\label{g50}
\end{figure*}

\begin{figure*} 
\hfill{}
\psfig{file=fig3.ps,width=7.5cm,%
bbllx=205pt,bblly=220pt,bburx=437pt,bbury=620pt,clip=}
\hfill{}
\caption[]{Filtered large-scale total intensities of the area shown in
Fig.~2. Contours start at 4500~mK~T$_\mathrm{B}$ and run in steps of
50~mK~T$_\mathrm{B}$. The beam width of the filter was $3\degr$.}
\label{g50b}
\end{figure*}

\begin{figure*} 
\hfill{}
\psfig{file=fig4.ps,width=13.5cm,%
bbllx=80pt,bblly=75pt,bburx=510pt,bbury=760pt,clip=}
\hfill{}
\caption[]{Polarized intensities of the area shown in Fig.~2. Contours
start at 0~mK~T$_\mathrm{B}$ and run in steps of 60~mK~T$_\mathrm{B}$.
The wedge shows the lower and upper cuts of the image. This field is
an example of a typical medium latitude region as the structured
emission features decrease with increasing latitude.}
\label{g50c}
\end{figure*}

\begin{figure*} 
\begin{minipage}{10cm}
\hfill{}
\psfig{file=fig5a.ps,width=9.7cm,%
bbllx=180pt,bblly=60pt,bburx=510pt,bbury=760pt,clip=}
\end{minipage}
\begin{minipage}{10cm}
\psfig{file=fig5b.ps,width=9.7cm,%
bbllx=180pt,bblly=60pt,bburx=510pt,bbury=760pt,clip=}
\hfill{}
\end{minipage}
\caption[]{The left panel shows the total intensity map towards the
 northern part of the  Cygnus region.
The three contour sets start at 0~mK~T$_\mathrm{B}$, 800~mK~T$_\mathrm{B}$
and 1800~mK~T$_\mathrm{B}$ (white contours) and run in steps of
150~mK~T$_\mathrm{B}$, 300~mK~T$_\mathrm{B}$ and 750~mK~T$_\mathrm{B}$,
respectively.
The right panel shows the total intensity map towards the southern
part of the Cygnus region.
The contours start at 0~mK~T$_\mathrm{B}$ and run in steps of
150~mK~T$_\mathrm{B}$. The contours plotted in white start at
1200~mK~T$_\mathrm{B}$ and run in steps of 400~mK~T$_\mathrm{B}$.
In both of the panels the electric field vectors are scaled to the polarized
intensity and  100~mK~T$_\mathrm{B}$ represented with a bar of length
$8\arcmin$. Every third vector is plotted.}
\label{north_tp}
\end{figure*}

\begin{figure*} 
\begin{minipage}{10cm}
\hfill{}
\psfig{file=fig6a.ps,width=9.7cm,%
bbllx=180pt,bblly=60pt,bburx=510pt,bbury=760pt,clip=}
\hfill{}
\end{minipage}
\begin{minipage}{10cm}
\hfill{}
\psfig{file=fig6b.ps,width=9.7cm,%
bbllx=180pt,bblly=60pt,bburx=510pt,bbury=760pt,clip=}
\hfill{}
\end{minipage}
\caption[]{Large-scale total intensity map towards the northern 
part of the  Cygnus region is shown in the left panel.
The first contour set starts at 
4160~mK~T$_\mathrm{B}$ and run in steps of 50~mK~T$_\mathrm{B}$ and 
the second contour set starts at 4800~mK~T$_\mathrm{B}$ and run in 
steps of 300~mK~T$_\mathrm{B}$. Cyg~A at about $\ell \sim 76\degr$ is blanked.
The right panel displays the large-scale total intensity map towards the 
southern part of the Cygnus region. The contour set starts at 
4300~mK~T$_\mathrm{B}$ and run in steps of 50~mK~T$_\mathrm{B}$}
\label{north_bg}
\end{figure*}

\begin{figure*} 
\begin{minipage}{10cm}
\hfill{}
\psfig{file=fig7a.ps,width=9.7cm,%
bbllx=180pt,bblly=60pt,bburx=510pt,bbury=760pt,clip= }
\hfill{}
\end{minipage}
\begin{minipage}{10cm}
\hfill{}
\psfig{file=fig7b.ps,width=9.7cm,%
bbllx=180pt,bblly=60pt,bburx=510pt,bbury=760pt,clip= }
\hfill{}
\end{minipage}
\caption[]{
The right panel shows the polarized intensity map 
towards the northern part of the  Cygnus region.
Contours start at 50~mK~T$_\mathrm{B}$ and run in steps of
50~mK~T$_\mathrm{B}$. Cyg~A at about $\ell \sim 76\degr$ is blanked.
Polarized intensity map towards the  southern part of the Cygnus region
is given in the left panel.
Contours start at 30~mK~T$_\mathrm{B}$ and run in steps of
40~mK~T$_\mathrm{B}$.
}
\label{north_pi}
\end{figure*}

\begin{figure*} 
\hfill{}
\psfig{file=fig8.ps,width=13.5cm,%
bbllx=80pt,bblly=70pt,bburx=510pt,bbury=760pt,clip= }
\hfill{}
\caption[]{Total intensity map close to  $\ell ~\sim 140\degr$.
Contours start at 0~mK~T$_\mathrm{B}$ and run in steps of
100~mK~T$_\mathrm{B}$. Small-scale polarization data are overlaid as
vectors such that 100~mK~T$_\mathrm{B}$ corresponding to a bar of
length $6\arcmin$. Every second vector is plotted.}
\label{g146}
\end{figure*}

\begin{figure*} 
\hfill{}
\psfig{file=fig9.ps,width=16cm,%
bbllx=52pt,bblly=267pt,bburx=585pt,bbury=516pt,clip= }
\hfill{}
\caption[]{Large-scale total intensity map near $\ell \sim
140\degr$. Contours start at 4530~mK~T$_\mathrm{B}$ and run in steps of
10~mK~T$_\mathrm{B}$. Large-scale electric field vectors are also
overlaid. A vector whose length is $6\arcmin$ corresponds to an
intensity of 100~mK~T$_\mathrm{B}$.}
\label{g146back}
\end{figure*}

\begin{figure*} 
\hfill{}
\psfig{file=fig10.ps,width=13.5cm,%
bbllx=80pt,bblly=60pt,bburx=510pt,bbury=760pt,clip= }
\hfill{}
\caption[]{Polarized intensity map of the small-scale emission near 
$\ell \sim 140\degr$. Contours start at 0~mK~T$_\mathrm{B}$ and run in
steps of 50~mK~T$_\mathrm{B}$. This map partly covers the field from
which the highest polarization emission in the Galaxy is observed.}
\label{g146_pi}
\end{figure*}

\begin{figure*} 
\hfill{}
\psfig{file=fig11.ps,width=16cm,%
bbllx=52pt,bblly=267pt,bburx=585pt,bbury=516pt,clip= }
\hfill{}
\caption[]{Large-scale polarized intensity map near $\ell \sim
140\degr$. Contours start at 380~mK~T$_\mathrm{B}$ and run in steps of
15~mK~T$_\mathrm{B}$.}
\label{g146back2}
\end{figure*}

\begin{figure*} 
\hfill{}
\psfig{file=fig12.ps,width=14.5cm,%
bbllx=80pt,bblly=60pt,bburx=510pt,bbury=760pt,clip=}
\hfill{}
\caption[]{Total intensity intensity map  of the small-scale emission
in the direction of the Galactic anticentre.
 Contours start at 0~mK~T$_\mathrm{B}$ and
run in steps of 50~mK~T$_\mathrm{B}$. Overlaid bars are electric field
vectors such that 100~mK~T$_\mathrm{B}$ corresponds to $4\arcmin$.
Every second vector is plotted. }
\label{antitp}
\end{figure*}

\begin{figure*} 
\hfill{}
\psfig{file=fig13a.ps,width=18.0cm,%
bbllx=100pt,bblly=50pt,bburx=510pt,bbury=760pt,rotate=90,clip=}
\hfill{}

\hfill{}
\psfig {file=fig13b.ps,width=18.0cm,%
bbllx=100pt,bblly=50pt,bburx=510pt,bbury=760pt,rotate=90,clip=}
\hfill{}
\caption[]{Total intensity map of the small-scale emission (at top) and the
polarized intensity map in the direction of  the Galactic anticentre. Upper and lower cuts of the images are shown with a wedge. }
\label{antitp_col}
\end{figure*}

\begin{figure*} 
\hfill{}
\psfig{file=fig14.ps,width=16cm,%
bbllx=51pt,bblly=251pt,bburx=585pt,bbury=530pt,clip= }
\hfill{}
\caption[]{Large-scale total intensity map in the direction of the
Galactic anticentre. Contours start at 3800~mK~T$_\mathrm{B}$ and 
run in steps of 25~mK~T$_\mathrm{B}$. }
\label{antiback}
\end{figure*}

\subsection{The area centered on $\ell = 50\degr,\ b = 12\degr$}

Fig.~\ref{g50} shows the small-scale total intensity emission with
superimposed polarization vectors in E-field direction.
Fig.~\ref{g50b} shows the corresponding large-scale emission
in total intensity and polarized intensities are shown in
Fig.~\ref{g50c}.

This field of $160~\sq$ is located
between the North Polar Spur to the west and the Cygnus-X region to the
east, which both contain strong emission from local sources. The
observed field seems less affected by local features and thus more
representative for the medium latitude emission from the inner part
of the Galaxy. The total intensity decreases smoothly with latitude
(Fig.~\ref{g50b}). Significant intensity variations in the small-scale
total intensity up to $20\degr$ latitude are visible in Fig.~\ref{g50}.
The polarized intensities (Fig.~\ref{g50c}) vary on small scales, but
become more uniform for latitudes above $16\degr$. A
chain of high-velocity \ion{H}{i}-clouds extending from
$\ell = 70\degr,\ b = 25\degr$ towards the Galactic plane terminates
close to $\ell = 52\degr,\ b = 10\degr$, where enhanced synchrotron
emission is visible as discussed by Uyan{\i}ker
(\cite{uyaniker97}).

\subsection{The Cygnus superbubble}

Fig.~\ref{north_tp} shows the small-scale
emission of the southern and northern part of the Cygnus-X area in the same
presentation as for Fig.~\ref{g50}. In 
Fig.~\ref{north_bg}  the corresponding large-scale total intensities
are given and polarized intensities are shown in 
Fig.~\ref{north_pi}.

The particular interest in this region arises from the
large X-ray halo surrounding the Cygnus-X area, called the
Cygnus superbubble (Cash et al.\ \cite{cash+80}). The Cygnus-X area is
a quite  strong and rather complex region in the radio range, since the
line of sight is along the local spiral arm. No sensitive radio surveys
of the area of the Cygnus superbubble exist so far and we will discuss
the relation of the X-ray emission revealed by ROSAT during its all-sky
 survey with the 1.4~GHz radio emission in
a forthcoming paper. Highly varying and
strong polarized intensity is seen in some areas, where the total
intensity images show rather smooth emission. In particular the regions
centered around $\ell = 75\degr,\ b = -12\degr$ south of the Cygnus loop
($\ell = 74\degr,\ b = -8\degr$), the large emission feature
almost filling the field of view and crossing the map diagonally  and
the structure at $\ell = 94\degr,\ b = -5\degr$ about $3\degr$ in size
 should be mentioned.

Some well-known polarized objects
are visible. These are three supernova remnants: the Cygnus
loop, HB21 ($\ell = 89\degr,\ b = 5\degr$) and W63 ($\ell = 82\degr,\
b = 5\fdg 5$). At  $\ell = 76\degr,\  b = 6\degr$ the
exceptionally strong emission from the radio galaxy Cygnus~A shows up.
Telescope sidelobes from the four support legs of the subreflector show 
up to about $4\degr$ distance from the
source in total intensity (Fig.~\ref{north_tp}) and slightly less
in polarized intensity (Fig.~\ref{north_pi}).


\subsection{The highly polarized region near $\ell = 140\degr$}

Fig.~\ref{g146} shows the area from $140\degr \leq \ell \leq 153\degr$
similar to Fig.~\ref{g50}, but polarization vectors represent the small-scale
emission component. To separate the small-scale polarization structures,
the strong sources are clipped in the absolutely calibrated  U and Q maps
and these maps are convolved to $3\degr$.
The convolved maps are subtracted from the original U and Q maps and
the polarized intensity maps are prepared.
 Fig.~\ref{g146back} is the corresponding large-scale
emission in total intensity with superimposed polarization vectors from
the large-scale component. Polarized intensities are shown in
Fig.~\ref{g146_pi} (small-scale emission) and Fig.~\ref{g146back2}
(large-scale emission), respectively.

Early polarization surveys as reviewed by Salter \& Brown
(\cite{salter+brown88}) already revealed an outstanding polarized region with
more than $20\degr$ in extent centered roughly at $\ell = 140\degr$ and
slightly north of the Galactic plane. Because of its morphology this area
has been referred to as the "fan region". The polarized emission is
believed to
be of local origin. The derived Rotation measures are small (e.g.
Bingham \& Shakeshaft\ \cite{bing+shake67}) and the
magnetic field direction has to be basically orientated
perpendicular to the line of sight. However, in addition to the
large-scale component, which we add from the Dwingeloo survey, a lot of
small-scale variations are visible (Fig.~\ref{g146_pi}).


\subsection{The anticentre region}

Fig.~\ref{antitp} to
Fig.~\ref{antiback} display the results of the anticentre region north of
the Galactic plane.
Color images (Fig.~\ref{antitp_col}) of the total intensity and
polarization intensity  in the direction of the Galactic
anticentre show the polarization features and absence of corresponding
total-power emission in detail.
Although an absolute calibration of the polarized emission on large
scales was not possible because of missing Dwingeloo data, significant
small-scale polarization is found across the area.
 The most remarkable
structures are apparently depolarized features (Fig.~\ref{antitp_col}),
which form filaments (e.g. at $\ell = 204\degr,\ b = 12\degr$ or
$\ell = 206\degr,\ b = 9\degr$) or ring-like structures (e.g. at $\ell
= 191\degr,\  b = 9\degr$ or at $\ell = 196\fdg 5,\ b = 5\fdg 5$)
with sizes up to about $3\degr$. The depolarized features have no
counterpart in the small-scale total intensity emission (Fig.~\ref{antitp},
~\ref{antitp_col}) or the large-scale component (Fig.~\ref{antiback}).
Depolarization can be caused either by some filamentary thermal matter
with enhanced electron density, by magnetic field variations in
strength and/or direction or the superposition of magnetic field
components in the line of sight with orientations perpendicular to each
other. Higher frequency observations with the 100-m telescope
are underway to clarify the nature of these features.


\section{Concluding remarks} \label{remarks}

The maps shown include information on a wealth of faint emission
structures at medium Galactic latitudes. The total intensities
decrease quickly with increasing distance from
the plane as is expected for a thin disk of emission. Strong sources
become rare, but faint extended ridges, arcs and more or less
structured discrete emission regions are seen superimposed on the still
dominating unresolved Galactic emission.

There is substantial diffuse polarized emission seen in
all the observed fields. This emission is characterized by fluctuations
on angular scales of the beam size up to several degrees.
Interestingly, there exist significantly polarized regions of
the order of several degrees in size, which have no corresponding
structures in the total-power emission. Moreover, within the
polarized emission there are numerous nearly straight, loop or arc
shaped structures, which seem to be depolarized. They are most
pronounced in the direction of  the Galactic anticentre, where the line of sight
across the Galaxy is comparably short. In these regions, the total
intensity of diffuse Galactic emission as revealed from the Stockert
1.4~GHz survey is less than 0.5~K. In view of the observed
polarized intensities of up to 0.25~K,  there are also areas with well
organized magnetic fields and quite small Faraday rotation effects.
The most likely explanation for the small-scale structures
are significant spatial variations of the Faraday rotation by the
interstellar medium in the line of sight, which modulate a
significantly polarized smooth Galactic diffuse emission. It is not
clear whether fluctuations of the Galactic magnetic field
or changes in the electron density are the reason for that.
We conclude that probing the properties of the diffuse
Galactic emission via a polarization survey is a promising tool towards
the improvement of our current understanding of the Galaxy and the
local ISM.


\end{document}